# Machine learning models with different cheminformatics data sets to forecast the power conversion efficiency of organic solar cells

Omar A. Álvarez-Gonzaga[1], Ulises A. Vergara-Beltran[1], Juan I. Rodríguez [2,*]

[1] Escuela Superior de Física y Matemáticas, Instituto Politécnico Nacional, Edificio 9, San Pedro Zacatenco, Ciudad de México C.P. 07738, México

[2] Centro de Investigación en Ciencia Aplicada y Tecnología Avanzada, Unidad Querétaro, Instituto Politécnico Nacional, Querétaro C.P. 76090, México

**ABSTRACT**

Random Forest (RF) and Gradient Boosting Regression Trees (GBRT) regression models along with three cheminformatics data sets (RDkit, Mordred, Morgan) have been used to predict the power conversion efficiency (PCE) of organic solar cells (OSCs). The data consists of cheinformatics descriptors of the electron donor used in 433 OSCs for which the experimental PCE (target variable) is reported in the literature. The donor is either a polymer or a small organic molecule, and the acceptor the fullerene derivatives PCBM or $PC_{71}BM$. Unlike previous methods, our ML approach considers the type of donor and the acceptor by adding four extra donor's features using the one-hot encoder tool. It is demonstrated that this additional information improves the prediction performance up to 34%. We have also exploited this feature to theoretically forecast the PCE of new OSCs by evaluating the ML model for a different acceptor. It is predicted that more than 50% of the OCSs obtained by exchanging the acceptor would have higher experimental PCE. The prediction accuracy of a given ML approach is analyzed for different PCE intervals. RF using RDkit descriptors resulted in the best ML approach with a Pearson's correlation coefficient for the training and testing sets equal to 0.96 and 0.62, respectively.

[*] **Corresponding Author**: jirodriguezh@ipn.mx

# I. Introduction

Solar cells (SC) have represented a clean and renewable energy source for about 70 years. However, this technology has not been used at a massive scale due mainly to its high cost, relatively low efficiency, and the requirement of large flat areas for solar exposure. Organic solar cells (OSCs) could potentially overthrow all OSs limitations. Made mainly from plastic (polymers), OSCs have the potential to be cheap, transparent, and flexible enough to be installed on a variety of non-conventional surfaces (windows, cars, walls, asphalt roads, etc.). Some OSC prototypes have reached a power conversion efficiency (PCE) of about $19\%$, which is already in the range of the PCE of some commercial SC's. The research on developing new cost-effective polymers and small organic molecules for producing OCSs with high PCE is ongoing worldwide. [1-4]

Predicting OSC's PCE by computer simulations is neither a simple nor quick venture. It is a multifactor and multiscale problem which is computationally demanding. For instance, in the so-called bulk-heterojunction (BHJ) OSC, the open circuit voltage ($V_{oc}$), a quantity that is directly proportional to the OSC's PCE, depends on several factors like the type of the active layer's donor-acceptor pair, the morphology of the active layer mainly at the donor-acceptor interface, relative donor-acceptor concentration, charge separation (*CS*) rate at the donor-acceptor interface ($K_{SC}$), exciton mobility and recombination, etc.. [6] Computing just $K_{SC}$ at the quantum mechanical level, based on density functional theory (DFT)-based calculations using Marcus theory, requires calculating several molecular parameters like the reorganizatión energy, the Gibbs free energy, and the electronic coupling of the CS process. [6-7] The BHJ-OSC active layer is an amorphous-like blend of donor and acceptors, which is typically modeled as a system with few donor (usually semiconductor polymers) and acceptor (usually fullerenes) molecules. However, the model system is relatively large, and the nature of the calculations involved demands the proper modeling of the electronic correlations to describe the realistic donor-acceptor interface morphology and the charge transfer process. [8-10] Calculating $K_{SC}$ for a single OSC requires both ground state (DFT) and excited state (TDDFT) calculations that could take several weeks or months of computer calculations due to the system's size and CPU time scaling. [7] Therefore, if computer simulations are considered tools for performing systematic massive donor-acceptor pair screening, faster and more efficient theoretical frameworks must be used. To this end, machine learning (ML) represents an intelligent option.

Recently, ML has been used to predict PCE and other properties of OSCs. ML regression and classification methods have been used to calculate OSCs PCE, giving different prediction performances. [11] Overall, ML methods have used DFT-based electronic properties (frontier orbitals, ionization potential, electron affinity, reorganization energy, etc.), [11-15] cheminformatics descriptors [16-18] or both of them [19] as the set of ML features. Using DFT-based descriptors usually implies solving the electronic structure problem for the ground an excited states for all systems in the training and validation sets using the standard CPU-high-cost methods. ML methods with cheminformatics descriptors are quicker to compute using many of the available specialized software. [11,16-19] Like in drug discovery, cheminformatics descriptors within the ML framework would make it possible to perform massive screening for the computer design of high-efficient cost-effective OSCs.

In this work, we have explored three types of cheminformatic data sets, RDkit, Mordred, and Morgan, in conjunction with RF and GBRT regression models to predict the OSC's PCE. A total of 433 OSCs, for which the experimental PCE is available, have been considered in the data set (see Figure 1). The prediction accuracy for each method and data set is analyzed. Experimentally, it is well known that the type of donor-acceptor pair impacts strongly on the PCE. An advantage of our ML methodology over previous works is that our ML approach considers the type of donor and the acceptor in the active layer of the OCSs by using the one-hot encoder tool. It is demonstrated that this extra information improves the prediction performance of the ML models. We have also exploited this feature to theoretically forecast the PCE of new OSCs by evaluating the ML model for a different acceptor.

## II. Methodology and Computational Details

Random Forest (RF) and Gradient Boosting Regression Trees (GBRT) regression methods, as implemented in the scikit-learn library, [20] were used to predict OSCs' PCE. The data set is composed of the information of 433 donors used in either BHJ or bilayer OCSs, which was obtained from references [15,19]. (see Supporting Information (SI)). This data was split into 90% and 10% for the training and the validation sets, respectively. The hyperparameters of each ML model were optimized by a 10-fold cross-validation process. We have computed the root-mean-square error (RMSE) and the Pearson's correlation coefficient ($r$) as metrics. The data was chosen so that the electron donor and acceptor could be of two types. The donor can be either a polymer or a small organic molecule; and the acceptor can be either the [6,6]-phenyl-C61-butyric acid methyl ester (PCBM) or the [6,6]-Phenyl $C_{71}$ butyric acid methyl ester ($PC_{71}BM$) (see Figure 1(a)-(c)). The side chains of the donor polymer were replaced with methyl groups to uniformize and simplify the calculations. It is known that these side chains do not have a direct impact on the PCE, but they are usually considered for solubility purposes. [21] In our ML approach, the type of donor and the acceptor information are considered as four extra donor features using a one-hot encoder tool, which assigns the value 1 or 0 if the corresponding entity (type of donor or acceptor) is present or not. From now on, we will label this information as "extra information" (EI). We have performed the training and validation processes with and without considering the EI to quantify the impact of the type of donor/acceptor on the particular ML model studied in this work. Experimentally, it is well known that the type of donor-acceptor pair impacts strongly on the PCE. [1] However, the previous ML approaches have not considered this information. [11-19] The experimental PCE and the Simplified Molecular Input Line Entry System (SMILES) [22] of each donor were obtained from references [15,19](see Figure 1). The SMILES were used to generate the cheminformatics descriptors for the three data sets considered here using the RDkit python library. [23] For a donor polymer, the SMILES of the corresponding monomer is considered. The number of RDkit, Mordred, [24] and Morgan [25] descriptors are 209, 1426, and 2048, respectively (see SI). The Morgan descriptors are arranged into a so-called fingerprint, a 2048-bit vector (see Figure 1(d)). [25]

## III. Results and Discussion

Figure 2 and Table 1 show the performance of all combinations of ML models and data sets for predicting the PCE. The metric values overall are in the range of those reported previously for ML models that use cheminformatic descriptors. [11-16-18] The correlation coefficient $r$ is greater than 0.9 for the training set for all ML methods, with an average equal to 0.94 (0.95) when(without) considering the extra information. The RMSE is lower than 1.01 for all ML methods, with an average equal to 0.78(0.72) when(without) considering the extra information. This indicates that the PCE is predicted with an error of about 1 unit, that is, 1% of PCE. These metric values are excellent for the training process. For the testing process, there is an overall decrease of r, which is in the interval [0.46, 0.64] ([0.51, 0.62]) when(without) considering the extra information. Overall, there is an improvement in the prediction(testing) when the EI of the type of donor/acceptor is considered in the ML model. Notice that the Pearson's r coefficient increases for each ML model-data set except for RF-Morgan when the EI is considered in the ML model. The greatest improvement (34.2%) is for RF-RDkit; the average r increase is 12.1%. Thus, as in many ML applications, using appropriate experimental prior-knowledge, like the donor-acceptor type in this particular application, makes the ML methods more effective. The best ML model is RF when trained with RDkit data, as seen clearly in Table 1. However, the other ML approaches show performance that is close to the RF-RDkit method. As remarked before, it is important to differentiate the prediction performance for different intervals of the target value (PCE) in this particular ML application. Certainly, it is more important that a given ML approach has better predictions at high PCE than at low PCE. The 2D histograms of Figure 3 supply information in this respect. In these 2D histograms, notice that for a ML model to have high prediction accuracy, the darkest squares (the most populated PCE intervals) should be close to the x-axis (low absolute error of predicted PCE versus reference PCE) and close to the right end (OSCs with high PCE). Thus, considering the training set, the darkest squares close to the x-axis indicate that most (over 95%) of the predicted PCE have an absolute error (AE) lower than 1%. This fact agrees with the excellent values of the metrics ( r and RMSE) for the training set (see Table 1). Overall, the darkest squares are not at higher PCEs not because of a drawback of the ML model but because of the original distribution of the PCE population, in which, unfortunately, the most populated intervals are below 5% of PCE (see Figure 1(a)). Having the 2D histograms for the two cases where the ML models are trained with and without considering the EI, detailed quantitative information on the ML model PCE prediction can be obtained for different PCE intervals. As a model example, let us consider the particular case of the RF-RDkit for the testing set(Figure 3(a)). For this case, the extra information has a three-fold positive effect. For the testing set with no EI, the interval with the most significant population (*4*) is for PCE in the interval 0-1% and an AE equal to $\approx 2-2.5\%$ . Considering the extra information, the most populated interval (5) undergoes a swift to higher PCE(3-4%) and lower absolute error ( $Error_{PCE} \approx 0-0.5\%$ ). It means that the extra information makes the RF-RDkit model's prediction accuracy notoriously increased for the OSCs with higher PCEs. Similar trends can be seen for GBRT-Mordred. However, this effect depends on the particular ML model and data set. For instance, for the GBRT-RDkit model, although these positive effects remain, they are diminished. For the GBRT-RDkit model , the number of OSCs with AE lower than 1% increased by only two units; the AE of the OSCs with PCE in the interval *7-8%* was lowered by only one unit. Overall, when considering the EI, this positive trend holds for all the ML approaches. The donor-acceptor type information not only improves the ML prediction performance but also allows the ML model to theoretically (computationally) design new OCSs by considering other combinations of donor and acceptors. Once the ML model is trained, it can be used to "evaluate" a different acceptor for a given donor and predict the PCE for the new donor-acceptor pair. We have performed this for twenty donors chosen at random whose PCE is greater than 5% (Figure 4(a)) and for the complete data set(training and validation) (Figure 4(b)). The RF-RDkit model (blue line in Figure 4(a)) recovers the reference's local and global trends (green line), which is an excellent result. Notice that the set of donor-acceptor pairs used in Figure 4 contains both training and test points. The PCE of new theoretically predicted OCSs (red line) is obtained by simply evaluating the ML model with the other different acceptor type. That is, if a given donor (label D1) in the data set has originally the PCBM as the acceptor, then the same donor (D1) is used with $PC_{71}BM$ as the acceptor in the ML model to predict the PCE of the OSC when using the D1- $PC_{71}BM$ pair in its active layer (and vice versa). For instance, the RF-RDkit model predicts that none of the new twentyOSCs of Figure 4(a) with the new donor-acceptor pair has a higher PCE than the OSC with the original pair. However, the PCE for several OSCs is increased upon acceptor exchange when all data is considered, as seen in Figure 4(b). In fact, the RF-RDkit model predicts that more than 50% of the OCSs obtained by exchanging the acceptor would have higher experimental PCE. The highest picks (red line local maxima between 200 and 300 in the x-axis) are particularly important. These picks indicate the OCSs with the most significant PCE increase upon acceptor exchange. Thus, these OCSs are the best candidates to be experimentally constructed.

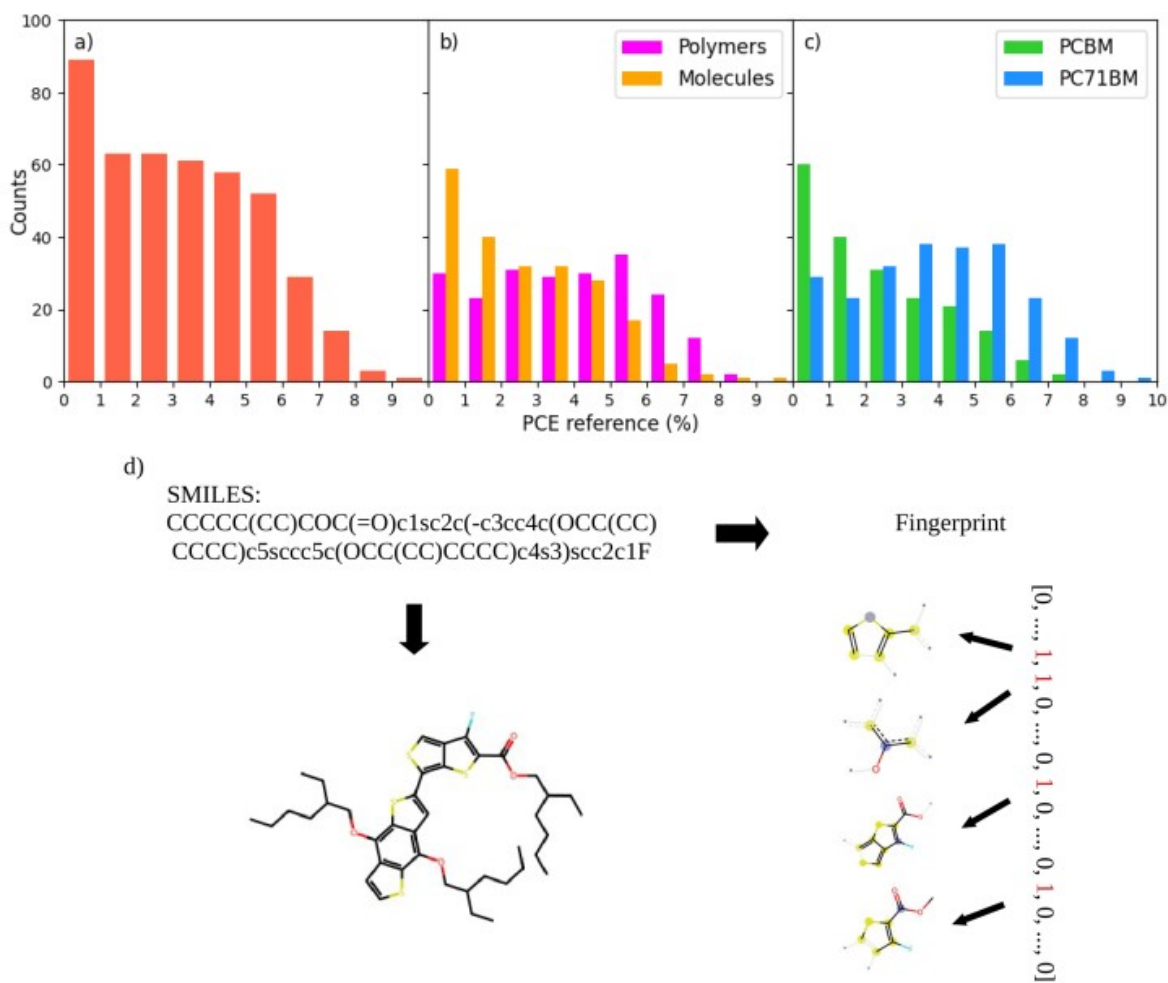

**Figure 1**: PCE histograms for the 433 OSCs considered in the data (a). The relative population of the OSCs with different type of donors (b) and acceptors (c). The SMILES for PTB7, its corresponding chemical diagram, and its Morgan fingerprint are shown in (d) as an example.

| Dataset | Model | Original information | | | | Extended information | | | |
|---|---|---|---|---|---|---|---|---|---|
| | | Training | | Testing | | Training | | Testing | |
| | | r | RMSE | r | RMSE | r | RMSE | r | RMSE |
| RDKit | RF | 0.957 | 0.783 | 0.467 | 1.877 | 0.968 | 0.691 | 0.627 | 1.675 |
| | GBRT | 0.951 | 0.745 | 0.482 | 1.857 | 0.956 | 0.701 | 0.515 | 1.829 |
| Mordred | RF | 0.972 | 0.644 | 0.596 | 1.702 | 0.974 | 0.628 | 0.599 | 1.701 |
| | GBRT | 0.980 | 0.489 | 0.566 | 1.765 | 0.981 | 0.484 | 0.609 | 1.685 |
| Morgan | RF | 0.903 | 1.012 | 0.641 | 1.632 | 0.930 | 0.862 | 0.598 | 1.703 |
| | GBRT | 0.905 | 1.005 | 0.542 | 1.795 | 0.914 | 0.951 | 0.604 | 1.703 |

**Table 1**: Pearson's correlation coefficient (r) and the root-mean square error (RMSE) for the original (without EI) and extended (with EI) datasets for all ML models.

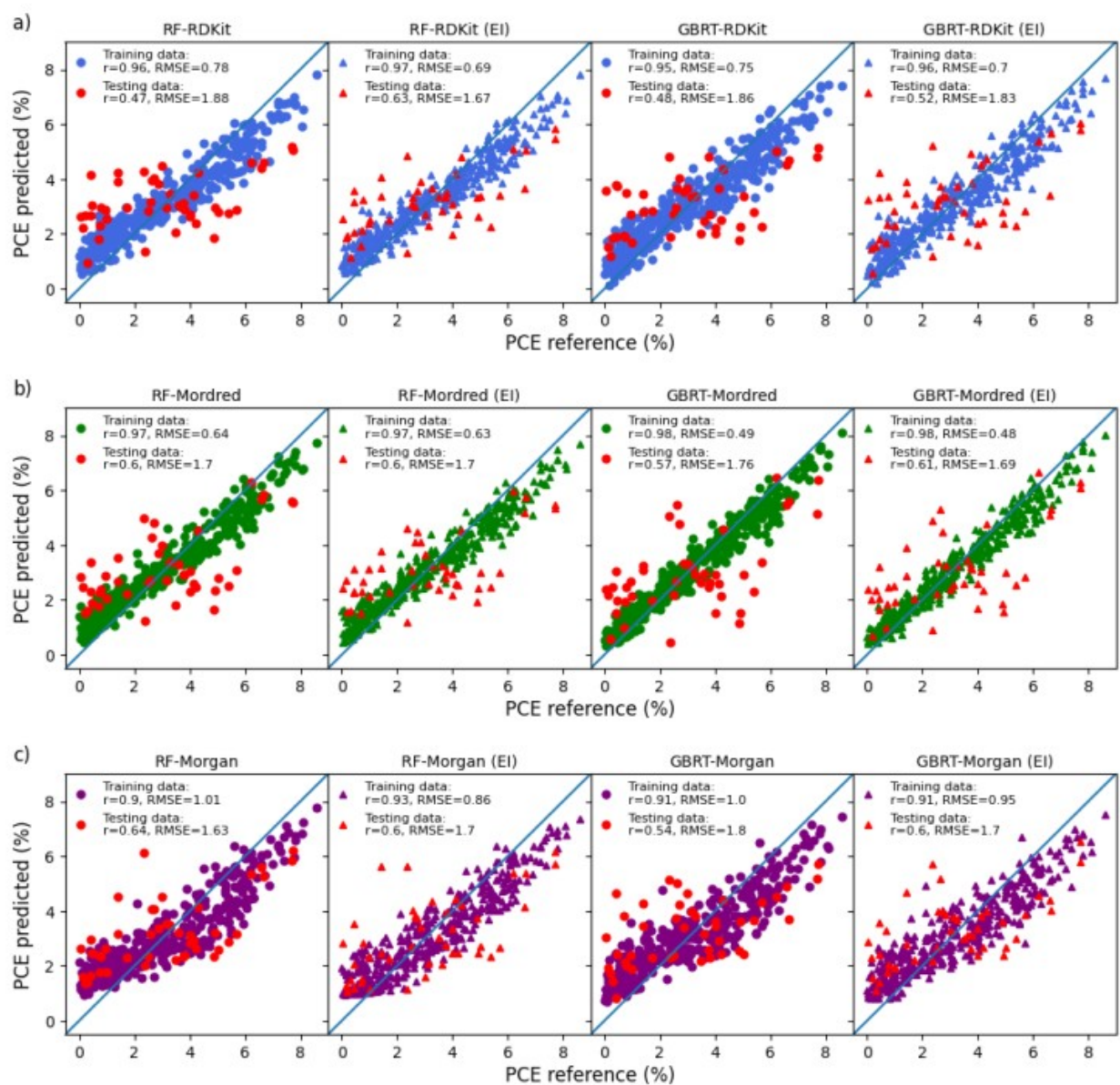

**Figure 2**: ML predicted vs experimental (reference) PCE for all combinations of ML model function and cheminfomatics original (without EI) and extended (with EI) sets. The metrics for both the training and testing sets are shown is the inset.

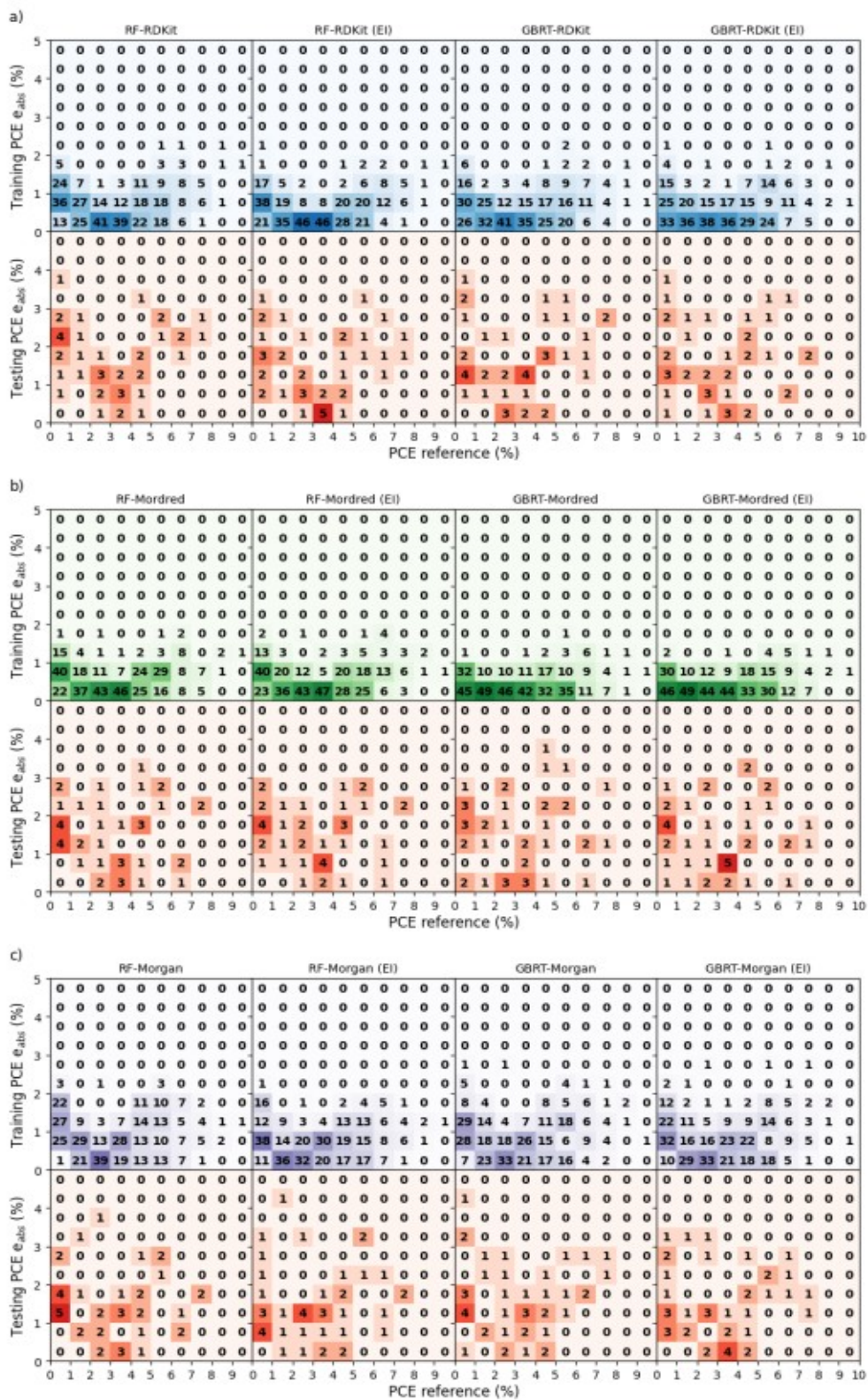

**Figure 3**: 2D-histograms for all combinations of ML models and cheminformatics data sets for the trainign (blue) and testing (red) sets. The color goes from light (low population) to dark (high population), the actual population is the number inside the squre. The reference PCE and the absolute error (PCEeabs) between it the ML predicted PCE is plotted on the x-axis and y-axis, respectively.

## IV. Conclusions

In this work, we have presented a comparative study of different ML approaches. The prediction performance of two ML regression (RF a GBRT) methods along three cheminformatics data sets (RDkit, Mordred, Morgan) have been analyzed. The information related to the type of both the electron donor and acceptor has been taken into account in the ML model as an extra four donor features by using the one-hot encoder tool. It was demonstrated that this extra information improves the prediction accuracy of the ML method. When using the extra information, the most significant improvement is for the RF-RDkit since the correlation coefficient r is increased by 34.2%. Overall, the average increase of r is 12.1%. Thus, as in many ML applications, using appropriate experimental prior knowledge, like the donor-acceptor type in this case, makes the ML methods more effective. Overall, RF using RDkit descriptors resulted in the best ML approach with a Pearson correlation coefficient $r_{training}=0.96$ and $r_{testing}=0.62$ for the training and testing sets, respectively. The RF-RDkit model predicts that more than 50% of the OCSs obtained by exchanging the acceptor would have higher experimental PCE.

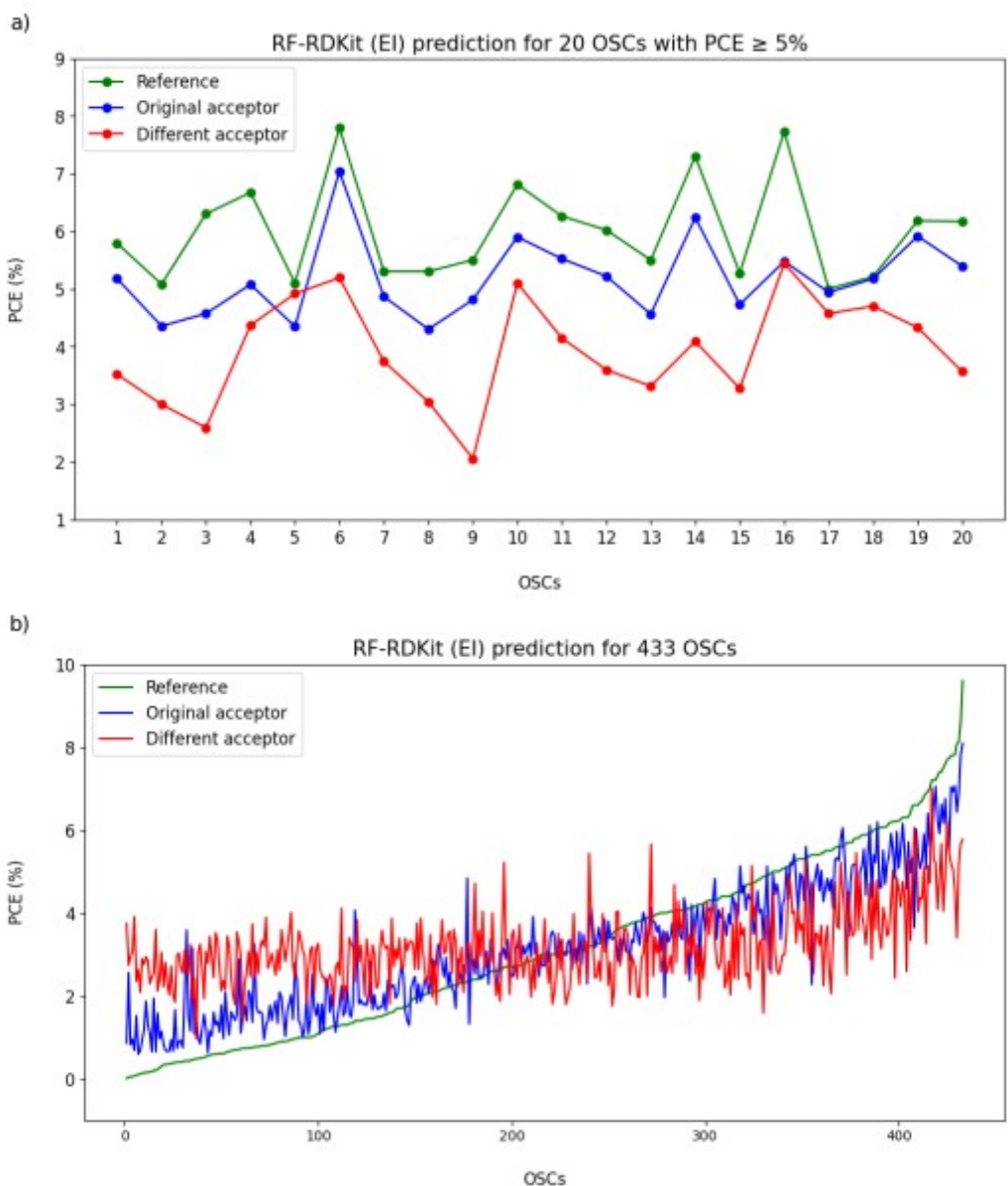

**Figure 4**: The green line shows the reference (experimental) PCEs. The blue line shows the ML-predicted PCE with the RF-RDkit model. The red line shows the theoretically forecasted PCE (obtained using the RF-RDkit model) for potential OSCs with the same electron donor (lines green and blue) but different from the original (experimental) acceptor. (a) Twenty OSCs with PCE > 5% chosen at random. (b) All OSCs in the data set. In (b), the OSC's donor were relabeled in increasing order.

**ACKNOWLEDGMENTS**

The authors acknowledge the "Secretaría de Investigación y Posgrado", Instituto Politécnico Nacional-México, for financial support (project SIP-20240537). O.A.A.G. thanks CONAHCYT-Mexico for the doctorate fellowship.

119:e25883

10. Gutiérrez-González I, Rodríguez JI, Molina-Brito B, Götz AW, Castillo-Alvarado FL (2014) Structural and electronic properties of the P3HT-PCBM dimer: a theoretical study. Chem Phys Lett 612:234–239

11. Yaping Wen and Haibo Ma. Accelerated discovery of new molecules for excitonic solar cells via machine learning and virtual screening. Chem. Modell., 2021, 16, 1–38.

12. Harikrishna Sahu, Weining Rao, Alessandro Troisi,* and Haibo Ma. Toward Predicting Efficiency of Organic Solar Cells via Machine Learning and Improved Descriptors. Adv. Energy Mater. 2018, 1801032.

13. Nastaran Meftahi et al.. Machine learning property prediction for organic photovoltaic devices. npj Computational Materials (2020) 166.

14. Steven A. Lopez, Benjamin Sanchez-Lengeling, Julio de Goes Soares, and Alán Aspuru-Guzik. Design Principles and Top Non-Fullerene Acceptor Candidates for Organic Photovoltaics. oule 1, 857–870, December 20, 2017.
15. A. Kuzmich, D. Padula, H. Ma and A. Troisi, Energy Environ. Sci., 2017, 10, 395.

16. W. Sun, Y. Zheng, K. Yang, Q. Zhang, A. A. Shah, Z. Wu, Y. Sun, L. Feng, D. Chen, Z. Xiao, S. Lu, Y. Li and K. Sun, Sci. Adv., 2019, 5, eaay4275.

17. S. Nagasawa, E. Al-Naamani and A. Saeki, J. Phys. Chem. Lett., 2018, 9, 2639.

18.  F.-C. Chen, Int. J. Polym. Sci., 2019, 4538514.

19. D. Padula, J. D. Simpson and A. Troisi, Mater. Horiz., 2019, 6, 343.

20. Pedregosa *et al.*, Scikit-learn: Machine Learning in Python, JMLR 12, pp. 2825-2830, 2011.

21. Liu, T.; Troisi, A. Absolute Rate of Charge Separation and Recombination in a Molecular Model of the P3HT/PCBM Interface. J. Phys. Chem. C 2011, 115, 2406−2415.

22. Anderson E, Veith GD, Weininger D (1987). SMILES: A line notation and computerized interpreter for chemical structures (PDF). Duluth, MN: U.S. EPA, Environmental Research Laboratory-Duluth. Report No. EPA/600/M-87/021.

23. RDKit: Open-source cheminformatics. https://www.rdkit.org

24. Moriwaki, H., Tian, YS., Kawashita, N. *et al.* Mordred: a molecular descriptor calculator. *J Cheminform* **10**, 4 (2018).

25. Rogers, D.; Hahn, M. Extended-Connectivity Fingerprints. J. Chem. Inf. Model. 2010, 50 (5), 742−754.